\documentclass[a4paper]{jpconf}
\usepackage{graphicx}
\begin{document}
\title{Searches for Higgs bosons with dark matter \\ at the Large Hadron Collider}
\author{Michele Gallinaro$^{1,2}$}
\address{$^1$on behalf of the ATLAS and CMS collaborations}      
\address{$^2$Laborat\'orio de Instrumenta\c{c}\~ao e F\'isica Experimental de Part\'iculas, LIP Lisbon, Portugal}

\ead{michgall@cern.ch}

\begin{abstract}
Convincing and direct evidence for dark matter (DM) on galactic scales comes from the observation of the rotation curves of galaxies. 
At particle colliders, searches for DM involve the production of a pair of stable electrically neutral and weakly interacting particles 
with a signature of missing transverse energy ($E^{\rm T}_{\rm miss}$) recoiling against a SM particle. 
The resulting signature yields a final state denoted as X+$E^{\rm T}_{\rm miss}$, where the SM particle X is emitted as initial state radiation.
The Higgs boson discovery at the LHC opens a new window into the searches for new physics processes 
beyond the SM through the h+$E^{\rm T}_{\rm miss}$ signature, as a direct probe of the interaction involving DM particles.
Due to the small Yukawa couplings to quarks and gluons, the initial state radiation of the Higgs boson is suppressed, 
but it can be produced in the case of a new interaction with DM particles.
Searches for DM particles produced in association with the Higgs boson are discussed. 
They are based on proton-proton collision data at the LHC in different final states.
\end{abstract}


\section{Introduction}
The Higgs boson discovered in 2012 by the ATLAS~\cite{Aad:2012tfa} and CMS~\cite{Chatrchyan:2012xdj} collaborations has properties consistent with those expected in the standard model (SM).
Its discovery marks the triumph of the SM and completes its self-consistency. 
However, despite the beautiful design and many experimental confirmations, the SM remains an incomplete description of Nature. 
The SM answers many of the questions about the structure of matter, but there are still several unanswered questions. Why do we observe matter and almost no anti-matter if we believe there is a symmetry between the two in the Universe?
Are quarks and leptons actually fundamental, or made up of even more fundamental particles?
Why are there three generations of quarks and leptons? What is the explanation for the observed pattern for particle masses?
How does gravity fit into all of this?
Although most of the matter in the Universe is dark matter (DM), its underlying particle nature remains unknown and cannot be explained within the SM.
What is this DM that we cannot see but has gravitational effects in the cosmos?

There is strong astrophysical and cosmological evidence that suggests for the existence of DM 
and that it makes up approximately 26\% of the total mass of the Universe~\cite{Bertone:2004pz,Ade:2015xua}. 
The evidence is based on its gravitational interaction. There is no evidence yet for non-gravitational interactions between DM and SM particles.
A number of BSM theories predict the particle origin of DM and several types of particle candidate are proposed in these models. Some of the most popular models propose the DM in the form of stable, electrically neutral, weakly interacting massive particle (WIMPs)~\cite{Steigman:1984ac} with a mass in the range between a few GeV to a few TeV, which opens up the possibility of searches at a particle collider.

A search for DM at a collider involves the need to look for a recoil against visible SM particles.
Due to the lack of electric charge and weak interaction cross section, the probability that DM particles produced in proton-proton collisions interact with the detectors~\cite{atlas,cms}
is expected to be very small,
and can be sought via an imbalance in the total momentum transverse to the beams, seen in the detector. 
Thus, many searches for DM at the LHC involve missing transverse momentum ($E^{\rm T}_{\rm miss}$) 
where a SM particle, X,
is produced against the missing transverse momentum, associated with the DM particles escaping the detector, in the so called ``$E^{\rm T}_{\rm miss}$+X'' or ``mono-X'' final states.
In the searches performed at colliders, X may be a  jet, a heavy flavor jet, a photon, 
or a W or Z boson~\cite{Aad:2015zva,Khachatryan:2014rra,Khachatryan:2014uma,Aad:2014vea,Aad:2014tda,Chatrchyan:2012tea,Aad:2013oja,Sirunyan:2017hci}.

In this context, the discovery of the Higgs boson opens up new possibilities in searches for physics beyond the SM (BSM), complementing other mono-X searches.
If DM is indeed associated with the scale of electroweak symmetry breaking, signatures related to the Higgs boson are a natural place to search for it.
The Higgs boson could be produced from a new interaction between DM and the SM particles~\cite{Carpenter:2013xra,Petrov:2013nia}.
The mono-Higgs signature can be produced as a result of final-state radiation of DM particles, or in a case of a new interaction of DM particles with the Higgs boson, via a
mediator particle. In all cases, the DM particle is denoted by $\chi$ and may be a fermion or a scalar.
There is an important difference between mono-Higgs and other mono-X searches. In proton-proton collisions, a jet/photon/W/Z can be emitted directly from a light quark as initial state radiation (ISR) through the usual SM gauge interactions, or it may be emitted as part of the new effective vertex coupling DM to the SM. 
In contrast, since Higgs boson ISR is highly suppressed due to the small coupling of the Higgs boson to quarks, a mono-Higgs is preferentially
produced through a direct coupling with dark matter particles. This can happen either via a decay of the boson into invisible particles, or via associated production with DM particles. 
Therefore, invisible decays of the Higgs boson or its production in association with $E^{\rm T}_{\rm miss}$ can be sensitive probes
of the effective DM-SM coupling, and of the structure of the BSM physics responsible for producing DM. 

Both the ATLAS and CMS collaborations have searched for such topologies. 
Events in the mono-X signatures are selected by requiring large $E^{\rm T}_{\rm miss}$ together with a SM particle for triggering purposes. 
The dominant backgrounds come from W/Z+jets and from irreducible di-boson and multijet processes.
In order to discriminate the signal against the backgrounds one has to understand the transverse momentum ($p_{\rm T}$) 
spectrum of the SM particles in order to correctly model the $E^{\rm T}_{\rm miss}$ distribution. 
By constraining the backgrounds in dedicated control regions (CRs) and then performing a simultaneous fit of CRs and signal region (SR), it is possible to extract the results.
In the mono-Higgs searches, different SM Higgs boson decay modes in events with missing transverse momentum are considered.
Mono-Higgs studies generally follow two general paths, either the effective field theory (EFT) approach that does not specify the underlying physics and is suppressed by powers of 1/$\Lambda$ where $\Lambda$ is the effective mass scale, or simplified models where explicit models are considered. The EFT approach involves direct couplings between DM particles and the Higgs boson, it is more model-independent but is not reliable when parton energies are comparable to the energy scale $\Lambda$, whereas simplified models are more model-dependent. The approaches are therefore complementary and both are explored.

\section{Invisible Higgs decays}

Invisible Higgs boson decays provide a way for exploring possible DM-to-Higgs boson couplings, provided that such decays are kinematically allowed. Null results from searches at the LHC for an invisibly decaying Higgs boson produced in association with a Z boson, combined with current Higgs boson data, already provide a model-independent constraint on the Higgs invisible branching ratio of B$_{\rm inv} <24$\% at 95\% C.L.~\cite{hinvisible}. 

In the SM, the Higgs boson decays invisibly only through the $h\rightarrow ZZ \rightarrow 4\nu$ process, with a small branching ratio of approximately B$\approx 0.1\%$. 
This rate may be significantly enhanced in the context of several BSM physics scenarios, including those in which the Higgs boson acts as a portal to DM.
The search for invisible decays 
of a Higgs boson, using proton-proton collision data was performed in several final states. 
The search targets events in which a Higgs boson is produced in association with jets from vector boson fusion (VBF)~\cite{Sirunyan:2018owy,Aaboud:2018sfi}.
This characteristic signature allows for the suppression of SM backgrounds.  The analysis exploits the kinematic features of the VBF topology by fitting the dijet invariant mass distribution. The dominant background (approximately 95\%) comes from V+jet processes.
Although it has a small cross section, the VBF channel is the most sensitive mode for invisible decays of a Higgs boson at hadron colliders.

Searches for invisible decays of the Higgs boson also target the associated production (Vh, where V denotes a W or Z boson)~\cite{Sirunyan:2017jix,Aaboud:2017bja,Sirunyan:2017qfc}
The Vh channels include both a search for Zh production, in which the Z boson decays leptonically, and one where a boosted W or Z boson decays hadronically and its decay products are captured by a single reconstructed large-radius jet. Additional sensitivity is achieved by including a search for gg$\rightarrow$gh production (hereafter referred to as ggH), where a high-$p_{\rm T}$ Higgs boson candidate is produced in association with initial-state jet radiation. 
The searches where the hadronic final states are present, e.g. ggh and V(hadronic)h, use the leading jet to trigger the event and have large cross section ($\approx 50$~pb). However, they are less sensitive due to the large backgrounds from $Z(\rightarrow \nu\nu)$+jets and $W(\rightarrow \ell\nu)$+jets (where the $\ell$ is not identified) but are nonetheless used to enhance the overall sensitivity. 
The results of the combination are further interpreted in the context of Higgs portal models of DM interactions.

\section{Higgs boson and dark matter}

Invisible Higgs boson decays are not sensitive to DM with a mass above $m_h/2 \approx 60$~GeV. Therefore, it is worthwhile 
to also investigate signatures with visible decays of the Higgs boson in association with invisible particles.
The signal events are characterized by the $E^{\rm T}_{\rm miss}$+Higgs signature, 
where the Higgs boson decay products recoil against the $E^{\rm T}_{\rm miss}$ and are used to trigger the event. 
The search is performed in five Higgs boson decay channels: a $b\bar{b}$ quark-antiquark pair, a pair of $\tau$ leptons, or a pair of photons, or W or Z bosons.
No significant excess over the expected SM background is observed in any of these five decay modes.

\subsubsection*{$\bf h\rightarrow b\bar b$:}
As the $h\rightarrow b\bar{b}$ decay mode has the largest branching ratio of all decay modes allowed in the SM, it provides the largest signal yield. 
However, it suffers from a very large background 
due to a large production of dijets as well as to a poor $E^{\rm T}_{\rm miss}$ resolution.
Backgrounds to the $E^{\rm T}_{\rm miss}$+$b\bar{b}$ final state include: $Zh$ production with $Z\rightarrow \nu\bar{\nu}$, an irreducible background, $Wh$ production with $W\rightarrow \ell\nu$ where the lepton from the W decay is not identified, $Zb\bar{b}$ and $Wb\bar{b}$.
A search for DM produced in association with a Higgs boson decaying to b-quarks~\cite{Aaboud:2017yqz,ATLAS:2018bvd,CMS:2018gbc} is performed using events with large $E^{\rm T}_{\rm miss}$ and either two b-tagged small-radius jets or a single large radius jet containing two b-tagged sub-jets. Jet identification is exploited by analyzing the jet substructure in order to optimize the search sensitivity over a broad mass range. The two b-jets from the Higgs boson decays can either be resolved or merged and divided according to exclusive energy regions.
A fit to the invariant mass of the Higgs boson candidate $m_{\rm h}$, based on a binned likelihood approach, is used to search for a signal, where $m_{\rm h}$ is represented by the dijet invariant mass of the two leading small-R jets in the resolved SR, and the leading large-R jet mass in the merged SR. The search is performed by looking for a bump in the invariant mass distribution in different $E^{\rm T}_{\rm miss}$ regions.

\subsubsection*{\bf Heavy resonances:}

Several SM extensions invoke massive gauge bosons ($W'$ and $Z'$) with weak couplings to the SM particles. 
Among these are the minimal $W'$ and $Z'$ models, strongly coupled composite Higgs models, and little Higgs models.
A large number of these models are described by the heavy vector triplet (HVT) framework which extends the SM by introducing a triplet of heavy vector bosons, one neutral ($Z'$) and two electrically charged ($W^{'\pm}$), which are degenerate in mass and are collectively referred to as $V'$.
A search for heavy resonances, decaying into the SM vector bosons and the SM Higgs boson, $V'\rightarrow Vh$, is performed~\cite{Sirunyan:2018qob}.
The Higgs boson is assumed to decay to a $b\bar{b}$ pair, and the vector boson V to decay to final states containing 0, 1, or 2 charged leptons,
depending on its decay mode ($Z\rightarrow\nu\nu$, $W\rightarrow\ell\nu$, $Z\rightarrow\ell\ell$), together with $E^{\rm T}_{\rm miss}$ due to the undetected neutrino(s).
In the $Z'$-2HDM model, the $Z\rightarrow\nu\nu$ decay is replaced by the pseudo-scalar $A$ decaying into DM particles, $A\rightarrow\chi\chi$. 
The signal would appear as a localized excess in the VH mass spectra above a smoothly falling distribution of the SM V+jets and $t\bar{t}$ backgrounds.
Events are divided into categories depending on the number and flavor of the reconstructed charged leptons.
In order to discriminate against the copious vector boson production, 
the identification criteria for the boosted $h\rightarrow b\bar{b}$ candidate is performed using the ``soft-drop'' jet mass algorithm~\cite{Sirunyan:2018qob},
that looks at the jet substructure to remove wide-angle soft radiation from a jet.
The invariant mass $m_{\rm Vh}$ or invariant transverse mass $m^{\rm T}_{\rm Vh}$ spectra are fit with a combined likelihood function and interpreted 
in the context of different models to extract the limits on the $V'$ production cross section.

\subsubsection*{$\bf h\rightarrow \gamma\gamma, \tau\tau$:}

Searches for DM particles are also performed by looking for events with large transverse momentum imbalance and a recoiling Higgs boson 
decaying to either a pair of photons or a pair of $\tau$ leptons~\cite{Aaboud:2017uak,Sirunyan:2018fpy}. 
Although the SM Higgs boson branching ratios to $\gamma\gamma$ and $\tau\tau$ are smaller than the branching fraction to $bb$, 
these two decay channels have unique advantages compared with the $h\rightarrow b\bar{b}$ channel. The $h \rightarrow \gamma\gamma$ channel benefits
from higher precision in reconstructed invariant mass, and the $h \rightarrow \tau\tau$ channel benefits from smaller SM background.
Additionally, the di-photon and di-tau final states are less dependent on $E^{\rm T}_{\rm miss}$ trigger thresholds, and the searches in these channels are complementary to
those in the $h\rightarrow b\bar{b}$ channel as they can probe DM scenarios with lower $E^{\rm T}_{\rm miss}$.
The branching fraction of the SM Higgs boson decaying into a pair of photons is small, B$\approx 0.2\%$, but
the final state has well measured objects, which leads to well measured $E^{\rm T}_{\rm miss}$ and the diphoton invariant mass system measured with a good energy resolution. 
Significant backgrounds to the $\gamma\gamma + E^{\rm T}_{\rm miss}$ final state include: $Zh$ production with $Z\rightarrow \nu\bar{\nu}$ is an irreducible background, 
or $Wh$ production with $W\rightarrow \ell\nu$ where the lepton is not identified, $Z\gamma\gamma$ with $Z\rightarrow \nu\bar{\nu}$.
The search in the $h \rightarrow \gamma\gamma$ channel uses a fit in the diphoton invariant mass spectrum to extract the signal yield.
In the $h \rightarrow \tau\tau$ channel, both leptonic and hadronic decays of the $\tau$ lepton are analyzed. After requiring an amount of $E^{\rm T}_{\rm miss}$ in order
to sufficiently suppress the QCD multijet background, the signal is extracted by performing a simultaneous fit to the transverse mass of the $E^{\rm T}_{\rm miss}$ and the two $\tau$
lepton candidates in the signal and control regions.

\subsubsection*{$\bf h\rightarrow WW$:}

The search in the $h\rightarrow WW$ decay channel is performed in the opposite-sign $e\mu$ final state that is less affected by the backgrounds. 
Due to the presence of the neutrinos preventing from reconstructing the Higgs mass peak, multi-variate analysis (MVA) techniques are used to build a discriminant.
The dominant background processes are from $t\bar{t}$ and non-resonant $WW$ production. W+jet production also contributes to the background when a jet is misidentified as a lepton.
Events are selected using single- and double-lepton triggers. Kinematical and topological selection cuts are applied to further enhance the signal significance. 
The invariant mass of the leptons from the signal events peaks at low values due to the presence of the two neutrinos, and the leptons are boosted along the same direction as the Higgs boson recoils against the DM particles.
A MVA approach that makes use of different kinematical variables is exploited to improve the separation between the signal and the main backgrounds, and maximize the sensitivity of the search.
A shape analysis with a likelihood fit of the resulting MVA discriminant is used to extract the cross section limits in the models considered.

\subsubsection*{$\bf h\rightarrow ZZ\rightarrow 4\ell$:}

The four-lepton decay mode, via $h\rightarrow ZZ\rightarrow 4\ell$ ($\ell=e,\mu$), has the smallest branching ratio of the main decay modes considered, but also offers the smallest backgrounds.
The main advantage of this over other $h$ decay modes is easily reducible backgrounds and a clean final state in which one can reconstruct the Higgs boson candidate.
The signal event topology is defined by the presence of four charged leptons ($4e$, $4\mu$, or $2e2\mu$) and $E^{\rm T}_{\rm miss}$ produced by the undetected DM particles.
Backgrounds to the $4\ell + E^{\rm T}_{\rm miss}$ final state include: $Zh$ production with $Z\rightarrow \nu\bar{\nu}$, an irreducible background,
$Zh$ production with $Z\rightarrow \ell\ell$ and $h \rightarrow \ell\ell\nu\nu$,
$Wh$ production with $W\rightarrow \ell\nu$ where the lepton from the $W$ decay is not identified, 
$h\rightarrow ZZ \rightarrow 4\ell$ or the continuum $(Z^{(*)}/\gamma^{*})(Z^{(*)}/\gamma^{*}) \rightarrow 4\ell$ production, with significant $E^{\rm T}_{\rm miss}$ values due to mis-measurement of leptons or soft radiation.
After applying the full event selection, the $E^{\rm T}_{\rm miss}$ distribution is used to compare data and the sum of the SM backgrounds. No excess over the expected number of background events is observed in data and exclusion limits on the cross section times the branching fraction, B$(h\rightarrow ZZ\rightarrow 4\ell)$, are set. 

\section{Long-lived particles}

Several possible extensions of the SM predict the existence of a dark sector that is weakly coupled to the visible one. Depending on the structure of the dark sector and its coupling to the SM, some unstable dark states may be produced at colliders and decay back to SM particles with sizeable branching fractions.
One case is 
when
a dark photon ($\gamma_d$) with mass in the MeV-to-GeV range mixes kinetically with the SM photon. 
Among the numerous models predicting $\gamma_d$, one class particularly interesting for the LHC features the hidden sector communicating with the SM through the Higgs portal.
If the dark photon is the lightest state in the dark sector, it will decay to SM particles, mainly to leptons and possibly light mesons. Due to its weak interactions with the SM, it can have a non-negligible lifetime.
At the LHC, these dark photons would typically be produced with a large boost due to their small mass, 
resulting in collimated jet-like structures containing pairs of leptons and/or light hadrons (Lepton-Jets, or LJs). 
In the benchmark models considered, the dark photons are produced from the decays of heavy particles and two Lepton-Jets are expected to be produced back-to-back in the azimuthal plane.
In the absence of a signal, the results of the search for LJ production are used to set upper limits on the product of cross section and Higgs decay 
branching fraction to LJs, as a function of the  $\gamma_d$ mean lifetime.
Assuming the SM gluon-fusion production cross section for a SM Higgs boson, its branching fraction to dark photons is found to be below 10\%~\cite{ATLAS:2016jza}.

\section{Summary}

Evidence for dark matter (DM) comes from astrophysical and cosmological observations and it is based on a gravitational interaction. However, nor gravitation nor DM are part of the SM.
The discovery of the Higgs boson at the LHC completes the standard model (SM). However, even with the inclusion of the Higgs boson, the SM is an incomplete description of Nature.
Its discovery opened new possibilities in searching for physics beyond the SM.
Searches for DM produced in association with Higgs bosons are discussed. Different decay modes of the Higgs boson are scrutinized in the searches. 
Events are characterized by a signature with missing transverse energy recoiling against SM particles from Higgs boson decay products or associated production.
Constraints on selected mediator-based dark matter models in a variety of final states are interpreted in terms of a set
of spin-1 and spin-0 single-mediator dark matter models~\cite{ATLAS_summary}.
No significant excess in the data over the expected SM backgrounds is observed, 
and upper limits on the production cross section as a function of the DM particle mass and branching fractions are set. 
Limits are complementary to those from direct DM search experiments.

\ack

To my friends and colleagues at the LHC who contributed to producing an incredible wealth of results in the past few years, 
to those who were crucial in making the CMS detector an incredible tool for precision measurements, and a far-sighted microscope for discovering the infinitesimally small.
To the teams that built and commissioned the LHC, and to those who control the accelerator complex at CERN, and eventually drive the proton beams into collision, 
because without such an outstanding performance those results would have not been possible.
A special thanks to the organizers of the ``Charged 2018'' workshop in Uppsala where -as usual- I found a welcoming atmosphere in a warm environment 
that allowed stimulating discussions and exchange of ideas.


\section*{References}
\begin{thebibliography}{9}

\bibitem{Aad:2012tfa} 
  G.~Aad {\it et al.} [ATLAS Collaboration],
  Phys.\ Lett.\ B {\bf 716}, 1 (2012).

\bibitem{Chatrchyan:2012xdj} 
S.~Chatrchyan {\it et al.}  [CMS Collaboration], 
Phys.\ Lett.\ B {\bf 716}, 30 (2012).

\bibitem{Bertone:2004pz} 
  G.~Bertone, D.~Hooper and J.~Silk,
  Phys.\ Rept.\  {\bf 405}, 279 (2005).
  
  \bibitem{Ade:2015xua} 
  P.~A.~R.~Ade {\it et al.} [Planck Collaboration],
  Astron.\ Astrophys.\  {\bf 594}, A13 (2016).
  
  \bibitem{Steigman:1984ac} 
  G.~Steigman and M.~S.~Turner,
  Nucl.\ Phys.\ B {\bf 253}, 375 (1985).


\bibitem{atlas}
  G.~Aad {\it et al.} [ATLAS Collaboration],
  JINST {\bf 3}, S08003 (2008).

\bibitem{cms}
  S.~Chatrchyan {\it et al.} [CMS Collaboration],
  JINST {\bf 3}, S08004 (2008).


  \bibitem{Aad:2015zva} 
  G.~Aad {\it et al.} [ATLAS Collaboration],
  Eur.\ Phys.\ J.\ C {\bf 75}, no. 7, 299 (2015);
  Erratum: [Eur.\ Phys.\ J.\ C {\bf 75}, no. 9, 408 (2015)].

\bibitem{Khachatryan:2014rra} 
  V.~Khachatryan {\it et al.} [CMS Collaboration],
  Eur.\ Phys.\ J.\ C {\bf 75}, no. 5, 235 (2015).

\bibitem{Khachatryan:2014uma} 
  V.~Khachatryan {\it et al.} [CMS Collaboration],
  Phys.\ Rev.\ Lett.\  {\bf 114}, no. 10, 101801 (2015).
  
  \bibitem{Aad:2014vea} 
  G.~Aad {\it et al.} [ATLAS Collaboration],
  Eur.\ Phys.\ J.\ C {\bf 75}, no. 2, 92 (2015).
  

\bibitem{Aad:2014tda} 
  G.~Aad {\it et al.} [ATLAS Collaboration],
  Phys.\ Rev.\ D {\bf 91}, no. 1, 012008 (2015);
  Erratum: [Phys.\ Rev.\ D {\bf 92}, no. 5, 059903 (2015)].
  
  \bibitem{Chatrchyan:2012tea} 
  S.~Chatrchyan {\it et al.} [CMS Collaboration],
  Phys.\ Rev.\ Lett.\  {\bf 108}, 261803 (2012).
  
  \bibitem{Aad:2013oja} 
  G.~Aad {\it et al.} [ATLAS Collaboration],
  Phys.\ Rev.\ Lett.\  {\bf 112}, no. 4, 041802 (2014).
  
  \bibitem{Sirunyan:2017hci} 
  A.~M.~Sirunyan {\it et al.} [CMS Collaboration],
  JHEP {\bf 1707}, 014 (2017).
  
  \bibitem{Carpenter:2013xra} 
  L.~Carpenter, A.~DiFranzo, M.~Mulhearn, C.~Shimmin, S.~Tulin and D.~Whiteson,
  Phys.\ Rev.\ D {\bf 89}, no. 7, 075017 (2014).
  
  \bibitem{Petrov:2013nia} 
  A.~A.~Petrov and W.~Shepherd,
  Phys.\ Lett.\ B {\bf 730}, 178 (2014).
  

  \bibitem{hinvisible}
  V.~Khachatryan {\it et al.} [CMS Collaboration],
    JHEP {\bf 02}, 135 (2017).

  \bibitem{Sirunyan:2018owy} 
  A.~M.~Sirunyan {\it et al.} [CMS Collaboration],
  arXiv:1809.05937 [hep-ex].
  
  \bibitem{Aaboud:2018sfi} 
  M.~Aaboud {\it et al.} [ATLAS Collaboration],
  arXiv:1809.06682 [hep-ex].
  
  \bibitem{Sirunyan:2017jix} 
  A.~M.~Sirunyan {\it et al.} [CMS Collaboration],
  Phys.\ Rev.\ D {\bf 97}, no. 9, 092005 (2018).
  
  \bibitem{Aaboud:2017bja} 
  M.~Aaboud {\it et al.} [ATLAS Collaboration],
  Phys.\ Lett.\ B {\bf 776}, 318 (2018).
  
  \bibitem{Sirunyan:2017qfc} 
  A.~M.~Sirunyan {\it et al.} [CMS Collaboration],
  Eur.\ Phys.\ J.\ C {\bf 78}, no. 4, 291 (2018).
  
  \bibitem{Aaboud:2017yqz} 
  M.~Aaboud {\it et al.} [ATLAS Collaboration],
  Phys.\ Rev.\ Lett.\  {\bf 119}, no. 18, 181804 (2017).
  
  \bibitem{ATLAS:2018bvd} 
  ATLAS Collaboration, 
  ATLAS-CONF-2018-039,
  http://cds.cern.ch/record/2632344.
  
  \bibitem{CMS:2018gbc} 
  A.~M.~Sirunyan {\it et al.} [CMS Collaboration],
  arXiv:1811.06562 [hep-ex].

  \bibitem{Sirunyan:2018qob} 
  A.~M.~Sirunyan {\it et al.} [CMS Collaboration],
  arXiv:1807.02826 [hep-ex].
  
  \bibitem{Aaboud:2017uak} 
  M.~Aaboud {\it et al.} [ATLAS Collaboration],
  Phys.\ Rev.\ D {\bf 96}, no. 11, 112004 (2017).
  
    \bibitem{Sirunyan:2018fpy} 
  A.~M.~Sirunyan {\it et al.} [CMS Collaboration],
  JHEP {\bf 09}, 046 (2018).
  
  

\bibitem{ATLAS:2016jza} 
  ATLAS Collaboration, 
  ATLAS-CONF-2016-042,
  http://cdsweb.cern.ch/record/2206083.
  
  \bibitem{ATLAS_summary} 
  ATLAS Collaboration, 
  ATLAS-CONF-2018-051,
  http://cds.cern.ch/record/2646248.
  
\end{thebibliography}
\end{document}